\documentclass{entcs}
\usepackage{entcsmacro}

\usepackage{amsmath}
\usepackage{amssymb}
\usepackage{stmaryrd}
\usepackage[all]{xy}
\usepackage{verbatim}
\usepackage{subfigure}

\newenvironment{nop}{}{}
\newsavebox{\codeboxBox}
\newcommand{\codeboxInput}[2][0.95]{%
\begin{lrbox}{\codeboxBox}%
\begin{minipage}{#1\textwidth}\scriptsize%
\verbatiminput{#2}%
\end{minipage}%
\end{lrbox}%
\par\noindent%
\begin{nop}\begin{center}%
\usebox{\codeboxBox}
\end{center}\end{nop}%
\par\noindent}

\newenvironment{codeboxInline}{%
\par\noindent\begin{nop}\footnotesize}{%
\end{nop}\par\noindent\ignorespacesafterend}

\newenvironment{sdisplaymath}{
\par\noindent\begin{nop}\small\begin{displaymath}}{
\end{displaymath}\end{nop}\ignorespacesafterend}

\newcommand{\minisection}[1]{%
{\vspace{1.5ex}\noindent\textbf{\textit{#1.}}~}}

\newcommand{\expr}[1]{\texttt{#1}}
\newcommand{\type}[1]{\texttt{#1}}
\newcommand{\abst}[2][]{\langle #2 \rangle_{A_{#1}}}
\newcommand{\conc}[2][]{\langle #2 \rangle_{C_{#1}}}

\def\lastname{Fluet and Pucella}
\begin{document}

\begin{frontmatter}
  \title{Practical Datatype Specializations with Phantom Types and Recursion Schemes} 
  \author{Matthew Fluet\thanksref{fluetemail}}
  \author{Riccardo Pucella\thanksref{riccardoemail}}
  \address{Department of Computer Science\\ 
           Cornell University\\ 
      Ithaca, NY 14853 USA}
  \thanks[fluetemail]
         {Email: \href{mailto:fluet@cs.cornell.edu} 
         {\texttt{\normalshape fluet@cs.cornell.edu}}}
  \thanks[riccardoemail]
         {Email: \href{mailto:riccardo@cs.cornell.edu} 
         {\texttt{\normalshape riccardo@cs.cornell.edu}}}
\begin{abstract} 
  Datatype specialization is a form of subtyping that captures program
  invariants on data structures that are expressed using the
  convenient and intuitive \texttt{datatype} notation.  Of particular
  interest are structural invariants such as well-formedness.  We
  investigate the use of phantom types for describing datatype
  specializations.  We show that it is possible to express
  statically-checked specializations within the type system of
  Standard ML.  We also show that this can be done in a way that does
  not lose useful programming facilities such as pattern matching in
  \texttt{case} expressions.
\end{abstract}
\end{frontmatter}

\section{Introduction}
\label{s:intro}

Data structures that are used pervasively in an application are often
implemented with a degree of genericity that makes them suited, though
not always well suited, to their various uses.  A benefit of this
genericity is that it ensures that core functionality of the data
structure is available to all clients.  A limitation is that a client
that is required to produce, consume, or maintain an instance of the
data structure subject to a particular invariant has difficulty
enforcing the invariant.  While many languages boast a type system
that statically enforces basic safety properties on data structures,
few allow programmers to directly capture these additional invariants
within the type system. This is unfortunate: when invariants are
reflected into a type system, compile-time type errors will indicate
code that could violate these invariants.

To make our discussion concrete, consider Boolean formulas, which we
use as a running example throughout this paper. A straightfoward
representation of formulas is the following:
\begin{codeboxInline}
\begin{verbatim}
  datatype fmla = Var of string | Not of fmla
                | True | And of fmla * fmla
                | False | Or of fmla * fmla
\end{verbatim}
\end{codeboxInline}
We can easily define a function \expr{eval} that takes a formula and
an environment associating every variable in the formula with a truth
value, and returns the truth value of the formula. Similarly, we can
define a \expr{toString} function that takes a formula and returns a
string representation of the formula. As is well known, a
propositional formula can always be represented in a special form
called Disjunctive Normal Form (or DNF), as a disjunction of
conjunctions of variables and negations of variables. A formula in DNF
is still a formula, but it has a restricted structure.  Some
algorithms that operate on formulas require their input to be
presented in DNF; therefore, it makes sense to statically check that a
formula is in DNF.

One way to perform this static checking is to simply introduce one
datatype for (generic) formulas and another datatype for DNF formulas,
and provide functions to explicitly convert between them. This is, of
course, inefficient. For example, converting a formula in DNF to a
string via \expr{toString} would require two complete traversals of
the DNF formula (one to convert it to a generic formula, and one to
build the string representation), as well as the allocation of an
intermediate structure (of the same size as the original formula). An
alternative is to define a DNF formula as a \emph{specialization} of
formulas.  For the sake of presentation, we assume a special syntax
for specializations. This syntax should be self-explanatory:
\begin{codeboxInline}
\begin{verbatim}
  datatype fmla   = Var of string | Not of fmla
                  | True | And of fmla * fmla
                  | False | Or of fmla * fmla
    withspec atom = Var of string
         and lit  = Var of string | Not of atom
         and conj = True | And of lit * conj
         and dnf  = False | Or of conj * dnf
\end{verbatim}
\end{codeboxInline}
Roughly speaking, the specialization \type{dnf} of the datatype
\type{fmla} is restricted so that the \expr{Or} constructor creates
list of conjunctions terminated with the \expr{False} constructor.  A
conjunction is defined by another specialization \type{conj} of the
datatype \type{fmla} that restricts the \expr{And} constructor to
forming lists of literals. A literal is essentially a variable or a
negated variable. This can be captured using two specializations,
\type{atom} for atomic literals and \type{lit} for literals. Notice
that to define the \type{dnf} specialization, we need all the
specializations \type{dnf}, \type{conj}, \type{lit}, and \type{atom}.
These specializations induce a simple subtyping hierarchy:
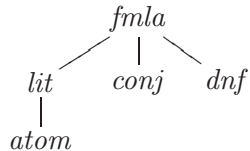
\begin{sdisplaymath}
\begin{xy}
\xymatrix@R=10pt@C=10pt{
& \mathit{fmla} \ar@{-}[dl] \ar@{-}[d] \ar@{-}[dr] & \\
\mathit{lit} \ar@{-}[d] & \mathit{conj} & \mathit{dnf} \\
\mathit{atom} 
}
\end{xy}
\end{sdisplaymath}
For uniformity, we consider the datatype as a degenerate
specialization.

Abstracting from this example, we define a \emph{specialization} of a
datatype to be a version of the datatype that is generated by a subset
of the datatype constructors, which may themselves be required to be
applied to specializations of the datatype.  If we view elements of a
datatype as data structures, then we can view elements of the
specialization as data structures obeying certain restrictions. A set
of specializations of a datatype may be specified in a mutually
recursive fashion.

In this paper, we show that we can implement, in Standard
ML~\cite{r:milner97}, much of what one would expect from a language
with a type system that directly supports the kind of specializations
described above.  For example, we can write a \expr{toDnf} function
that statically guarantees not only that its result is a formula, but
also that it is a DNF formula.  The advantage of implementing
specialization invariants in SML is that type systems directly
supporting specializations are complex and not widely available.  (A
type system that enforces similar, but strictly more powerful,
invariants is the \emph{refinement types}
system~\cite{r:davies05,r:freeman94}; see
Section~\ref{s:spec-pt-disc}.)

What are the key features of specializations that we would like
available?  For one, we would like the representation of values of
specialized types to be the same as the representation of the original
datatype.  This is important to avoid expensive run-time conversions
and to allow code reuse.  For example, we should be able to implement
a single function to evaluate not only an unspecialized formula, but
also any specialization of formulas, such as the \type{dnf}
specialization.  Moreover, we would like to write \expr{case}
expressions that do not include branches for constructors that do not
occur in the specialization of the value being examined.  For example,
if we perform a case analysis on a value with specialization
\type{dnf}, we should only need to supply branches for the
\expr{False} and \expr{Or} constructors.

Drawing on previous work relating phantom types and
subtyping~\cite{r:fluet02} and the intuition that the specializations
of a datatype induce a subtyping hierarchy, we present in
Section~\ref{s:spec-pt} an informal translation based on phantom types
from a set of specializations of a datatype to an interface providing
constructors, destructors, and coercions corresponding to the
specializations.  This interface forms a minimal set of primitive
operations that provide the functionality (and static guarantees) of
the specializations.  It uses the SML type system to enforce the
invariants embodied by the specializations.

Unfortunately, the interface leaves much to be desired, particularly
if we wish to promote datatype specializations as a practical
programming technique that integrates naturally with standard,
idiomatic SML usage.  In order to overcome these deficiencies, notably
the lack of pattern matching, we draw on another programming techique:
recursion schemes~\cite{r:wang03}.  We present in
Section~\ref{s:spec-rs} an informal translation from the set of
specializations of the datatype, using recursion schemes and the
interface described above, to an improved interface that provides the
functionality of the specializations through \emph{bona fide} SML
datatypes.  As before, the SML type system enforces the invariants of
the specializations.

The reader may well wonder why we choose to encode specializations in
SML, rather than designing a language extension.  First, there are
numerous SML
compilers~\cite{r:hamlet,r:mlkit,r:mlton,r:moscowml,r:polyml,r:sml.net,r:smlnj};
while many of these compilers implement some language extensions, the
SML language specified by the \emph{Definition}~\cite{r:milner97}
behaves the same in all compilers.  Hence, the technique described
here is available \emph{now} to \emph{all} SML programmers.
Furthermore, encoding specializations in SML is more expedient than
writing our own compiler or modifying an existing compiler.

Our notion of datatype specializations is quite general, and we
believe that it can capture a good number of useful invariants. For
example, we can define a specialization that ensures that the formula
contains no variables:
\begin{codeboxInline}
\begin{verbatim}
  datatype fmla   = Var of string | Not of fmla
                  | True | And of fmla * fmla
                  | False | Or of fmla * fmla
    withspec grnd = Not of grnd
                  | True | And of grnd * grnd
                  | False | Or of grnd * grnd
\end{verbatim}
\end{codeboxInline}
The following specializations distinguish between zero and non-zero:
\begin{codeboxInline}
\begin{verbatim}
  datatype nat       = Zero | Succ of nat
    withspec zero    = Zero
         and nonzero = Succ of nat
\end{verbatim}
\end{codeboxInline}
Lists give rise to interesting specializations. Consider the
following specializations, distinguishing between empty, singleton,
and nonempty lists:
\begin{codeboxInline}
\begin{verbatim}
  datatype 'a list        = Nil | Cons of 'a * 'a list
    withspec 'a empty     = Nil
         and 'a singleton = Cons of 'a * 'a empty
         and 'a nonempty  = Cons of 'a * 'a list
\end{verbatim}
\end{codeboxInline}
The following specializations distinguish between lists of even and
odd length:
\begin{codeboxInline}
\begin{verbatim}
  datatype 'a list   = Nil | Cons of 'a * 'a list
    withspec 'a even = Nil | Cons of 'a * 'a odd
         and 'a odd  = Cons of 'a * 'a even
\end{verbatim}
\end{codeboxInline}
Finally, we can use specializations to define abstract syntax trees
that distinguish between arbitrary expressions and well-formed
expressions (e.g., well-typed expressions
\cite{r:elliott00,r:leijen99}, expressions in normal forms, etc.). A
simple example of this is the following:
\begin{codeboxInline}
\begin{verbatim}
  datatype exp       = Bool of bool | And of exp * exp
                     | Int of int | Plus of exp * exp
                     | If of exp * exp * exp
    withspec boolexp = Bool of bool | And of boolexp * boolexp
                     | If of boolexp * boolexp * boolexp
         and intexp  = Int of int | Plus of intexp * intexp
                     | If of boolexp * intexp * intexp 
\end{verbatim}
\end{codeboxInline}
More involved examples that can be expressed using specializations
include red-black trees that check the critical invariant that no red
node has a red child after inserting a new element, and constructors
for expressions in the simply-typed $\lambda$-calculus that permit
only the building of type-correct expressions, essentially using an
encoding manipulating de Brujin indices.

\section{Specializations with Phantom Types}
\label{s:spec-pt}

How should we write (in SML) an implementation of the formula
specializations so that the type system enforces the appropriate
structural invariants? In this section, we give a highly-stylized
implementation that achieves this particular goal.  We hope that the
reader will grasp the straightforward generalization of this
implementation to arbitrary specializations.\footnote{Our running
example uses a first-order, monomorphic datatype, of which abstract
syntax trees are a typical example.  Extending the implementation to
handle first-order, polymorphic datatypes is straightforward.  It is
also possible to handle higher-order datatypes; we briefly consider
this in Section~\ref{s:spec-pt-disc}.}  We feel that a fully
elaborated example is more instructive than a formal translation,
where definitions and notation become burdensome and obfuscating.

We first review the essence of the phantom-types technique and its
application to subtyping~\cite{r:fluet02}.  The phantom-types
technique uses the definition of type equivalence to encode
information in a superfluous type variable of a type.  (Because
instantiations of this type variable do not contribute to the run-time
representation of values of the type, it is called a \emph{phantom type}.)
Unification can then be used to enforce a particular structure on the
information carried by two such types.

When applied to subtyping, the information we wish to encode is a
position within a subtyping hierarchy.  We require an encoding
$\langle\sigma\rangle$ of each specialization $\sigma$ in the
hierarchy; this encoding should yield a type in the SML type system,
with the property that $\langle\sigma_1\rangle$ unifies with
$\langle\sigma_2\rangle$ if and only if $\sigma_1$ is a subtype of
$\sigma_2$ in the hierarchy (written $\sigma_1\leq \sigma_2$).  An
obvious issue is that we want to use unification (a symmetric
relation) to capture subtyping (an asymmetric relation). The simplest
solution is to use two encodings $\conc{\cdot}$ and $\abst{\cdot}$
defined over all the specializations in the hierarchy. A \emph{value}
of specialization $\sigma$ will be given a type using $\conc{\sigma}$.
We call $\conc{\sigma}$ the \emph{concrete encoding} of $\sigma$, and
we assume that it uses only ground types (i.e., no type variables). In
order to restrict the domain of an operation or constructor to the set
of values that are subtypes of a specialization $\sigma$, we use
$\abst{\sigma}$, the \emph{abstract encoding} of $\sigma$.  In order
for the underlying type system to enforce the subtyping hierarchy, we
require the encodings $\conc{\cdot}$ and $\abst{\cdot}$ to
\emph{respect} the subtyping hierarchy by satisfying the following
property:
\begin{displaymath}
\mbox{
For all specializations $\sigma_1$ and $\sigma_2$, 
$\conc{\sigma_1}$ unifies with $\abst{\sigma_2}$ 
iff $\sigma_1\leq\sigma_2$}.
\end{displaymath}
To allow for unification, the abstract encoding introduces free
type variables. Since, in the SML type system, a top-level type cannot
contain free type variables, the abstract encoding is always a
part of some polymorphic type scheme.  This leads to some restrictions
on the uses of abstract encodings, the details of which are beyond the
scope of 
this paper (but, see our previous work~\cite{r:fluet02} for a thorough
discussion).  We assert that abstract encodings are used appropriately
in the following presentation.

\begin{figure}
\hrule
\codeboxInput{code/fmla/fmla.sig}%
\hrule
\caption{The \expr{FMLA} signature}
\label{f:fmlasig}
\end{figure}

\begin{figure}
\hrule
\codeboxInput{code/fmla/fmla.sml}%
\hrule
\caption{The \expr{Fmla} structure}
\label{f:fmlastruct}
\end{figure}

Figures~\ref{f:fmlasig} and \ref{f:fmlastruct} give a signature and
corresponding implementation of the specializations above. The amount
of code may seem staggering for such a small example, but we shall see
that most of it is boilerplate code.  (In fact, it is straightforward
to mechanically generate this code from a declarative description of
the specializations; see Section~\ref{s:concl}.) Moreover, all the
action is in the signature!  The implementation is trivial.  Part of
the reason for this explosion of code is that we implement
specializations notionally as abstract types with explicit
constructors and destructors.  (Section~\ref{s:spec-rs} shows how to
improve upon this seemingly draconian implementation.)  With this in
mind, let us examine the different elements of the signature and their
implementation.

\minisection{Types} 
The first part of the signature defines the types for the
specializations of formulas.  We introduce a polymorphic type \type{'a
t}, representing the values of the specializations. 

The first series of type abbreviations combines the abstract encodings
of the specializations with the specialization type, yielding the
\emph{abstract types}.  Consider the definition of the type
\type{ALit}:
\begin{codeboxInline}
\begin{verbatim}
  type 'a ALit = {lit: 'a} AFmla (* = {fmla: {lit: 'a}} t *)
\end{verbatim}
\end{codeboxInline}
Here, \verb+{fmla: {lit: 'a}}+ is the abstract encoding of the
specialization \type{lit} in the subtyping hierarchy given above. Note
that the sequence of record labels describes the path through the
subtyping hierarchy from the root to the \type{lit}
specialization.\footnote{This is essentially the encoding of tree
hierarchies given in our previous work~\cite{r:fluet02}.  All of the
other encodings given in that work are also applicable in this
setting.}

The second series of type abbreviations instantiates the abstract
types with \type{unit}, yielding the \emph{concrete types}.  We can
verify that \type{CLit} and \type{CAtom} unify with \type{ALit}. We
can also verify that all concrete types unify with \type{AFmla},
capturing the fact that the \type{fmla} specialization is the top
element of the subtyping hierarchy.  Yet \type{CLit} does not unify
with \type{AAtom}, \type{AConj}, or \type{ADnf}.  In other words, the
encodings respect the subtyping hierarchy.

In the implementation of Figure~\ref{f:fmlastruct}, we see that the
type \type{Rep.t} is implemented as a \emph{bona fide} SML datatype,
while the type \type{'a t} is implemented as a type abbreviation whose
polymorphic type variable is ignored, but serves as a placeholder for
a position in the subtyping hierarchy.  Hence, all the specializations
share the same representation.  We use \emph{opaque signature
matching} in Figure~\ref{f:fmlastruct}.  This is crucial to get the
required behavior for the phantom types.

\minisection{Constructors} 
For every specialization, the interface provides a function for each
constructor of the specialization. For instance, the \type{atom}
specialization has a single constructor (\expr{Var}), so we provide a
function\footnote{In the following and in
Section~\ref{s:spec-rs}, we occasionally replicate declarations in the
signature or structure using long identifiers where the SML syntax
requires an (unqualified) identifer.  We do so in order to
unambiguously denote the appropriate portion of the code; in all
cases, the meaning should be clear.}
\begin{codeboxInline}
\begin{verbatim}
  val Atom.Var : string -> CAtom
\end{verbatim}
\end{codeboxInline}
that returns an element of the specialization \type{atom} (and hence,
of type \type{CAtom}).  We allow subtyping on the constructor
arguments, where appropriate.  Hence, the \expr{And} constructor for
\type{conj} is available as:
\begin{codeboxInline}
\begin{verbatim}
  val Conj.And : 'a ALit * 'b AConj -> CConj
\end{verbatim}
\end{codeboxInline}

The implementation of these constructors is trivial. They are simply
aliases for the actual constructors of the representation type
(brought into scope by \expr{open Rep}).  Placing them in different
structures allows us to constrain their particular 
type, depending on the specialization we want them to yield.

\minisection{Destructors}
For every specialization, the interface provides a destructor function
that can be used to simultaneously discriminate and deconstruct
elements of the specialization, similar to the manner in which the
\expr{case} expression operates in SML.  Each destructor function
takes an element of the specialization as well as functions to be
applied to the arguments of the matched constructor.

As an example, consider \expr{Conj.dest}, the destructor function for
the specialization \type{conj}. Because the elements of \type{conj}
are built using only the \expr{True} and \expr{And} constructors,
deconstructing elements of such a specialization can only yield the
\expr{True} constructor or the \expr{And} constructor applied to a
\type{lit} value and to a \type{conj} value. Therefore, we give
\expr{Conj.dest} the type:
\begin{codeboxInline}
\begin{verbatim}
  val Conj.dest : 'a AConj -> {True : unit -> 'b, 
                               And : CLit * CConj -> 'b} -> 'b
\end{verbatim}
\end{codeboxInline}
Similar reasoning allows us to drop or refine the types of various
branches in the destructor functions for the other specializations.

Informally, the invariants of the specializations are preserved when
we use concrete encodings in covariant type positions and abstract
encodings in contravariant type positions.  This explains the
appearance of concrete encodings in the argument types of the branch
functions.

Destructor functions also have a trivial implementation. They are
simply implemented as an SML \expr{case} expression. On the branches
for which no function is provided, we raise an exception.  If the
invariants of the specializations are enforced, we know that those
exceptions will never be raised! By virtue of our encoding of
subtyping, static typing ensures that this exception will never be
raised by programs that use the interface~\cite{r:fluet02}.

\minisection{Coercions}
Finally, the interface provides coercion functions that convert
subtypes to supertypes.  Such coercion functions are necessary
because, intuitively, phantom types provide only a restricted form of
subtyping.  To illustrate the problem, consider the following
function; it does not typecheck because the type of the true branch is
\type{CLit} and the type of the false branch is \type{CAtom} -- two
types that cannot be unified:
\begin{codeboxInline}
\begin{verbatim}
  fun bad b = if b then Lit.Var ("p") else Atom.Var ("q")
\end{verbatim}
\end{codeboxInline}
Instead, we must write the following:
\begin{codeboxInline}
\begin{verbatim}
  fun good b = if b then Lit.Var ("p") else Lit.coerce (Atom.Var ("q"))
\end{verbatim}
\end{codeboxInline}
Technically, this behavior is due to the fact that type subsumption
occurs only at type application (implicit in SML), which most often
coincides with function application.  Thus, when two expressions of
different specializations occur in contexts that must have equal
types, such as the branches of an \expr{if} expression, subsumption
does not occur, and the expressions must be coerced to a common
supertype.  Coercions are also useful to work around a restriction in
SML that precludes the use of polymorphic
recursion~\cite{r:henglein93,r:kfoury93}.  We shall shortly see an
example where this use of a coercion is necessary.

The implementation of coercion functions is trivial. They are simply
identity functions that change the (phantom) type of a value.

\subsection{Examples} 
\label{s:spec-pt-ex}
Let us give a few examples of functions that can be written against
the interface of formula specializations given above.  First, consider
a simple function to identify the top-level operator of a formula:
\codeboxInput[0.90]{code/fmla/fmla-identify.sml}%
Note that the type of \texttt{identify}, \texttt{'a AFmla -> string},
asserts that the function may be safely applied to any formula
specialization.

A more interesting example is a recursive \expr{toString} function
that returns a string representation of a formula.  A simple
implementation is the following:
\codeboxInput[0.90]{code/fmla/fmla-toString.sml}%
Note that the inferred type of the \expr{toString'} function is
\type{CFmla -> string}, because it is recursively applied to variables
of type \type{CFmla} in the \expr{Fmla.dest} branches and SML does not
support polymorphic recursion.  However, we can recover a function
that allows subtyping on its argument by composing \expr{toString'}
with an explicit coercion.  Now the SML type system infers the desired
type for the \expr{toString} function.

If we had polymorphic recursion, we could directly assign the type
\type{'a AFmla -> string} to \expr{toString'}.  One may ask whether
the lack of polymorphic recursion in SML poses a significant problem
for the use of specializations as we have described them.
Fortunately, under reasonable assumptions,\footnote{The main
assumption is that the desired recursive function on the unspecialized
datatype can itself be written without polymorphic recursion.  If this
is not the case, then the function cannot be written in SML with or
without phantom types and specializations.  (Thus, we exclude
specializations of non-regular datatypes.)  On the other hand, if this
is the case, then one can write the function using only the interface
provided by specializations.} it can be shown that use of specializations never
needs polymorphic recursion in an essential way.  This is a
consequence of the fact that the use of phantom types in
specializations only influences the \emph{type} of an expression or
value, never its representation.  The argument proceeds as follows.

Consider a recursive function \expr{f} with type \type{$\sigma$ ->
$\tau$}, for a specialization $\sigma$.  Any recursive call in the
body of \expr{f} must be applied to an argument \expr{x} of
specialization $\sigma'$, where $\sigma'$ is a subtype of $\sigma$.
Since $\sigma'$ is a subtype of $\sigma$, there exists a coercion from
$\sigma'$ to $\sigma$.  Hence, we can always implement the function as
we did for the \expr{toString} function: set the domain of an auxilary function \expr{f'} to the
concrete encoding of $\sigma$ and write the recursive calls as
\expr{f' ($\sigma$.coerce x)}.  (In the \expr{toString} example, all of
the recursive calls are on values of type \type{CFmla}, so the
coercion can be elided.)  Finally, we recover the appropriate
subtyping in the argument of \expr{f} by composing \expr{f'} with
\expr{$\sigma$.coerce}.

\begin{figure}
\hrule
\codeboxInput{code/fmla/fmla-toDnf.sml}%
\hrule
\caption{The \expr{toDnf} function}
\label{f:todnf}
\end{figure}

Figure~\ref{f:todnf} gives an extended example culminating with a
\expr{toDnf} function that converts any formula into an equivalent DNF
formula.  The type of this function, \type{'a AFmla -> CDnf}, ensures
that the result formula is a DNF formula.  We further note that the
use of type annotations in Figure~\ref{f:todnf} is completely
superfluous.  Type inference will deduce precisely these types.

There is a difference between the guarantee made by the SML
type system and the guarantee made by a specialization library
employing the phantom-types technique.  While the former ensures that
``a well-typed program won't go wrong,'' the latter ensures that ``a
well-typed client won't go wrong, provided the specialization library
is correctly implemented.''  We hope that this section has
demonstrated that the implementation of a specialization library is
straightforward.  However, a more subtle point is that one must choose
specializations that capture properties and invariants of interest in
order to gain the benefits of additional static guarantees.

\section{Specializations with Recursion Schemes}
\label{s:spec-rs}

While Section~\ref{s:spec-pt} describes an interesting theoretical
result---the ability to define a \expr{toDnf} function whose type
statically enforces a structural invariant without recourse to
separate datatypes---it is not yet clear whether the methodology
presented is usable in practice.  In particular, pattern matching on
specialization types must be performed via application of the
destructor functions, rather than via SML's built-in syntactic support
for pattern matching.  The result is the ``truly unreadable
code''\footnote{courtesy of an anonymous reviewer} in
Figure~\ref{f:todnf}.  While any use of pattern matching can be
desugared to applications of the destructor functions, 
important aspects of the pattern-matching programming idiom are
seriously inhibited by the encoding in the previous section. 

Examining Figure~\ref{f:todnf} reveals two particularly glaring
assaults on readability that could be improved with pattern matching.
First, consider the \expr{andDnfs} function.  Informally, one can
describe the intended behavior of the function in the following way:
\textit{consider both arguments in the \type{dnf} specialization: if
either argument is a \expr{False} element, return \expr{False};
if both arguments are \expr{Or} elements, return the disjunction of
the pairwise conjunction of the elements' arguments.}  Unfortunately,
the written code obscures this behavior.  This highlights two missing
aspects of pattern matching: the ability to match simultaneously via
the nesting of datatype patterns within tuple patterns, and the
ability to give a wild-card match.

Second, consider the \expr{toDnf'} function, more specifically, 
the nested application of the \expr{caseFmla} destructor function.
Here, one misses the ability to write nested patterns, which would
combine the two applications of destructor functions into a single
pattern match.

\begin{figure}
\hrule
\codeboxInput{code/fmla/fmla-ns-toDnf.sml}%
\hrule
\caption{The \expr{toDnf} function (via the unspecialized \expr{Rep.t} datatype)}
\label{f:todnfns}
\end{figure}
For comparison, Figure~\ref{f:todnfns} implements the \expr{toDnf}
function for the unspecialized \expr{Rep.t} datatype, addressing the
concerns above.  However, this apparent improvement has come at a
cost.  The compiler now issues multiple nonexhaustive match warnings.
More importantly, the type system provides no assurance that the
result is in Disjunctive Normal Form.

We seek a solution that brings the expressiveness and
convienence of pattern matching to specializations.  Recall the
observation that we made in the introduction: the same static
invariants can be obtained by using distinct datatypes for each
specialization and providing functions to convert between them.  As we
pointed out, this is inefficient. However, there is a middle-ground
solution, one that uses distinct datatypes for their induced patterns
and fine-grained coercions to localize coercions to and from the
specialization types and the specialization datatypes.

\newsavebox{\footnotemarkBox}
\begin{lrbox}{\footnotemarkBox}{\footnotemark}\end{lrbox}
\begin{figure*}
\hrule
\codeboxInput{code/fmla/fmla-dt.paper.sig}%
\hrule
\caption{The \expr{FMLA\_DT} signature\usebox{\footnotemarkBox}}
\label{f:fmladtsig}
\end{figure*}

\begin{figure*}
\hrule
\codeboxInput{code/fmla/fmla-dt.sml}%
\hrule
\caption{The \expr{FmlaDT} structure}
\label{f:fmladtstruct}
\end{figure*}

We take as inspiration Wang and Murphy's recursion schemes
\cite{r:wang03}.  Exploiting two-level types, which split an
inductively-defined type into a component that represents the
structure of the type and a component that ties the recursive knot,
recursion schemes provide a programming idiom that can hide the
representation of an abstract type while still supporting pattern
matching.

Roughly speaking, the technique suggests defining a datatype for each
specialization, which represents the top-level structure of the
specialization.  Pattern matching on a specialization is performed by
first converting part of the specialization into the appropriate
datatype and then matching on the result.  The important point is that
we do not need to convert the whole specialization into the
specialization datatype, but rather only as much as is needed to
perform the pattern matching.

Figures~\ref{f:fmladtsig} and \ref{f:fmladtstruct} give a
signature.
and corresponding implementation of datatypes for the specializations
of the formulas.  (Again, while the quantity of code is large, it is
largely boilerplate code that can be mechanically generated.)
Note that the signature and implementation are written against the
\expr{FMLA} signature and \expr{Fmla} structure; in particular, the
\expr{FMLA\_DT} signature and \expr{FmlaDT} structure do not require
access to the representation type \type{Rep.t}.  Therefore, we need
not make any additional arguments about the safety of using the
datatype interface to the specializations.  That is, we cannot violate
the invariants imposed by the specializations by using the datatype
interface.  As we did in Section~\ref{s:spec-pt}, let us examine the
different elements of the signature and their implementation.

\footnotetext{The signature given for \expr{FMLA\_DT} is not valid
SML, in that the substructures (\expr{Fmla}, \expr{Lit}, etc.) 
are extended.  While this abuse of notation seems acceptable in an
exposition, an implementation must textually duplicate and extend the
\expr{FMLA} signature.  This argues that SML could benefit
from a more expressive language of signatures~\cite{r:ramsey05}.}

\minisection{Datatypes}
The first part of each substructure defines the datatype used to
represent a specialization.  The datatype for a specialization has a
datatype constructor for each constructor in the specialization.
Wherever a constructor has a specialization as an argument, we
introduce a type variable, creating a polymorphic datatype.  The
polymorphic type allows the datatype to represent unfoldings of the
specialization with encoded specialization types at the ``leaves'' of
the structure.  For instance, the \type{dnf} specialization
\begin{codeboxInline}
\begin{verbatim}
  withspec dnf = False | Or of conj * dnf
\end{verbatim}
\end{codeboxInline}
becomes
\begin{codeboxInline}
\begin{verbatim}
  datatype ('orl, 'orr) Dnf.t' = Dnf.False' | Dnf.Or' of 'orl * 'orr
\end{verbatim}
\end{codeboxInline}
replacing the references to the specializations \type{conj} and
\type{dnf} with the polymorphic type variables \type{'orl} and
\type{'orr}.  Thus, the structure of a \type{dnf} specialization is
given without reference to specific types for \type{conj} or
\type{dnf}.

The polymorphic type variables allow the specialization datatype to
represent arbitrary, finite unrollings of the specialization type.
For example, the type \type{(CConj, CDnf DDnf) DDnf} corresponds to
unrolling the \type{dnf} specialization into the \type{Dnf.t'} datatype
once at the top-level and once again at the second argument to the
\expr{Or} constructor.  Hence, it has as elements \expr{Dnf.False'},
\expr{Dnf.Or' (e1, Dnf.False')}, and 
\expr{Dnf.Or' (e1, Dnf.Or' (e2, e3))} for\linebreak any \expr{e1} and \expr{e2}
of type \type{CConj} and \expr{e3} of type \type{CDnf}.

\minisection{Injections}
For every specialization, the interface provides a function for
coercing from the specialization datatype to the specialization type.
In particular, the coercion is from one top-level unrolling of the
specialization type (into the specialization datatype) back to the
specialization type.  The implementation is straightforward: map each
datatype constructor to the corresponding specialization constructor.
For the \type{Dnf.t'} datatype, this yields
\begin{codeboxInline}
\begin{verbatim}
  fun Dnf.inj f = 
    case f of Dnf.False' => Dnf.False 
            | Dnf.Or' (f1, f2) => Dnf.Or (f1, f2)
\end{verbatim}
\end{codeboxInline}
with the type \type{('a AConj, 'b ADnf) DDnf -> CDnf}.  We find
injections to be used infrequently in practice.  One usually wishes to
build values at the specialization type, and the specialization
constructors are better suited for this purpose than injecting from
the specialization datatype.  Hence, we do not use injections in our
examples.

\minisection{Projections}
Much more practical are the projection functions, which convert from a
specialization type to the specialization datatype.  Again, the
conversion is from the specialization type to a single top-level
unrolling of the specialization type (into the specialization
datatype).  The implementation builds upon the destructor functions,
by mapping each dispatch function to the corresponding datatype
constructor.  For the \type{DDnf} datatype, this yields
\begin{codeboxInline}
\begin{verbatim}
  fun Dnf.prj f = 
    Dnf.dest f {False=fn () => Dnf.False', 
                Or = fn (f1,f2) => Dnf.Or' (f1,f2)}
\end{verbatim}
\end{codeboxInline}
with the type \type{'a ADnf -> (CConj, CDnf) DDnf}.

\minisection{Maps}
One final useful family of functions are the structure-preserving
maps.  These functions, similar in flavor to the familiar map on
polymorphic lists, apply a function to each polymorphic element of a
structure, but otherwise leave the structure's ``shape'' intact.  For
instance, the map
\begin{codeboxInline}
\begin{verbatim}
  fun Lit.map F f = 
    case f of Lit.Var' s => Lit.Var' s 
            | Lit.Not' f => Lit.Not' (F f)
\end{verbatim}
\end{codeboxInline}
transforms an \type{'not1 DLit} to an \type{'not2 DLit} via a function
\expr{F} of type \type{'not1 -> 'not2}.  Since the polymorphic
elements of a specialization datatype correspond to the nested
specializations used by that datatype, maps are useful for localizing
unfoldings of a specialization for nested pattern matching.

\subsection{Example}

\begin{figure*}
\hrule
\codeboxInput{code/fmla/fmla-dt-toDnf.sml}%
\hrule
\caption{The \expr{toDnf} function (via datatype interface)}
\label{f:todnfdt}
\end{figure*}

Figure~\ref{f:todnfdt} reproduces the code from Figure~\ref{f:todnf}
using the datatype interface to the formula specializations.  While
some may argue that the syntactic differences are minor, the
differences are important ones.  All of the functions are written in a
familiar, readable pattern-matching style.  In particular, the
\expr{andDnfs} function uses simultaneous pattern matching and
wild-card matches.  Also, the \expr{toDnf'} function uses nested
patterns to fold all the branches into a single \expr{case}
expression.  Note the use of the \expr{Fmla.map} function to unfold
only the \type{CFmla} under the \expr{Fmla.Not'} constructor, while
leaving all other \type{fmla} specializations folded.  The expression
discriminated by the \expr{case} has the following type:
\begin{codeboxInline}
\begin{verbatim}
  ((CFmla, CFmla, CFmla, CFmla, CFmla) DFmla, 
                  CFmla, CFmla, CFmla, CFmla) DFmla
\end{verbatim}
\end{codeboxInline}

Finally, a word about the efficiency of the compiled code.  In the
presence of cross module inlining, smart representation decisions, and
some local constant folding, the overhead of projections from a
specialization type to a specialization datatype can be almost
entirely eliminated.  Specifically, when projections appear directly
as the expression discriminated by a \expr{case}, then a compiler can
easily fold the \expr{case} expression ``buried'' in the projection
function into the outer \expr{case} expression.

\section{Discussion}
\label{s:spec-pt-disc}

We discuss the relationship between our implementation of
specializations and related approaches and techniques in the
literature.  We also discuss some limitations, and how they could be
lifted.

\minisection{Refinement Types} 
As we have already mentioned, refinement types~\cite{r:freeman94}
enforce invariants similar in spirit to (and strictly more expressive
than) specializations.  In fact, many of the examples considered here
were inspired by similar examples expressed using refinement types.
However, there are a number of critical differences between refinement
types and specializations as described and implemented in this paper.
These differences permit us to encode specializations directly in the
SML type system.  The most significant difference concerns the
``number'' of types assigned to a value.  In short, refinement types
use a limited form of type intersection to assign a value multiple
types, each corresponding to the evaluation of the value at specific
refinements, while our technique assigns every value exactly one
type.\footnote{Both systems in fact employ a form of subtyping to
further increase the ``number'' of types assigned to a value.}  For
example, in Section~\ref{s:spec-pt-ex}, we assigned the
\expr{litToDnf} function the (conceptual) type \type{lit -> dnf}.
With refinement types, it would be assigned the type \type{(fmla ->
fmla) $\land$ (lit -> dnf)}, indicating that in addition to mapping
literals to DNF formulae, the function can also be applied to an
arbitrary formula, although the resulting formula will not satisfy any
of the declared refinements.

While this demonstrates the expressiveness of refinement types, it
does not address the utility of this expressiveness.  In particular,
one rarely works in a context where \emph{all} possible typings of an
expression are necessary.  In fact, the common case, particularly with
data-structure invariants, is a context where exactly one type is of
interest: a ``good'' structure is either produced or preserved.  This
is exactly the situation that motivates our examples of
specializations.  In this sense, our technique is closer in spirit to
refinement-type checking~\cite{r:davies05}, which verifies that an
expression satisfies the user-specified refinement types.

\minisection{Deforestation}  
We pointed out in the introduction and again in
Section~\ref{s:spec-rs} that the same static invariants that we obtain
through the use of specializations can be obtained by using distinct
datatypes for each specialization and explicit conversion functions
between the datatypes. We also pointed out that this approach is
inefficient.  Deforestation is a compiler optimization that aims to
reduce the allocation of intermediate structures that are created when
using functions that operate on data structures such as lists and
trees~\cite{r:wadler90}. It is quite possible that a 
judicious use of such techniques can alleviate the inefficiency
inherent in using multiple datatypes for specialization.  While one
would gain more readable \texttt{case} expressions, avoiding the need
for projections, one would lose the subsumption of function arguments
available through the phantom-types technique.  Deforestation suffers
from the same criticism as refinement types: it is not widely
implemented in compilers, particularly in languages like SML, where
strict evaluation and effects often invalidate the optimization.  In
constrast, our approach is applicable independent of the underlying
compiler technology.

\minisection{Phantom Types} 
The phantom-types technique underlies many interesting uses of type
systems.  It has been used to derive early implementations of
extensible records~\cite{r:wand87,r:remy89,r:burton90}, to provide a
safe and flexible interface to the Network Socket
API~\cite{r:reppy96}, to interface with COM
components~\cite{r:finne99}, to type embedded compiler
expressions~\cite{r:leijen99,r:elliott00}, to record sets of effects in
type-and-effect type systems~\cite{r:pessaux99}, to embed a
representation of the C type system in SML~\cite{r:blume01}, to
guarantee conforming XHTML documents are produced by SML
scriplets~\cite{r:elsman04}, and to provide a safe interface to GUI
widgets~\cite{r:larsen04}.  The application to type-embedded compiler
expressions and to guarantee the production of conforming XHTML
documents are most closely related to our specialization technique.
Many of the other applications focus on enforcing a required
\emph{external} structure on an essentially unstructured
\emph{internal} representation (e.g., the SML representation of a
socket is just a 32-bit integer).  Our specialization technique
extends the idea to enforce \emph{internal} invariants on a common
\emph{internal} representation.

\minisection{Recursion Schemes}
While the implementation described in Section~\ref{s:spec-rs} is
inspired by recursion schemes~\cite{r:wang03}, there are significant
differences. Recursion schemes are designed to support monomorphic
recursive types, with a straightforward extension to polymorphic
recursive types.  However, it is less clear how to extend the idiom to
mutually recursive types, which may arise in specializations.  Hence,
we have chosen to allow each reference to a specialization to be typed
independently.  Clearly, we do not wish to identify \type{conj} and
\type{dnf} in the \type{dnf} specialization, because the two
specializations are distinct.  It is debatable whether all occurences
of the \type{fmla} specialization within the \type{fmla}
specialization should be identified; that is, whether we should
introduce
\begin{codeboxInline}
\begin{verbatim}
  datatype ('not,'andl,'andr,'orl,'orr) Fmla.t' = 
    Var' of string | Not' of 'not
  | True' | And' of 'andl * 'andr
  | False' | Or' of 'orll * 'orr
\end{verbatim}
\end{codeboxInline}
or
\begin{codeboxInline}
\begin{verbatim}
  datatype 'fmla Fmla.t' = 
    Var' of string | Not' of 'fmla
  | True' | And' of 'fmla * 'fmla
  | False' | Or' of 'fmla * 'fmla
\end{verbatim}
\end{codeboxInline}
The latter is closer to recursion schemes (and, hence, could be given
a simple categorical interface \cite{r:wang03}). For pragmatic
reasons, we have instead adopted the former, because it gives finer
grained control over coercions to and from the datatype.  However, as
our entire implementation does not require access to the
representation type, either or both definitions could be used without
difficulty. 

\minisection{Extensions} 
We have made a number of implicit and explicit restrictions to
simplify the treatment of specializations.  There are a number of ways
of relaxing these restrictions that result in more expressive systems.
For example, we can allow a specialization to use the same constructor
at multiple argument types.  Consider defining a DNF formula to be a
list of \expr{Or}-ed conjuctions that grows to either the left or the
right:
\begin{codeboxInline}
\begin{verbatim}
  withspec dnf = False | Or of conj * dnf | Or of dnf * conj
\end{verbatim}
\end{codeboxInline}
One can easily define two constructor functions for \expr{Or} that
inject into the \type{dnf} specialization.  However, the ``best''
destructor function that one can write is:
\begin{codeboxInline}
\begin{verbatim}
  val Dnf.dest : 'a ADnf -> {False : unit -> 'b, 
                             Or : CFmla * CFmla -> 'b} -> 'b
\end{verbatim}
\end{codeboxInline}
not, as might be expected,
\begin{codeboxInline}
\begin{verbatim}
  val Dnf.dest : 'a ADnf -> {False : unit -> 'b, 
                             Or : (CConj * CDnf -> 'b) * 
                                  (CDnf * CConj -> 'b)} -> 'b
\end{verbatim}
\end{codeboxInline}
While the second function can be written with the expected semantics,
it requires a run-time inspection of the arguments to the \expr{Or}
constructor to distinguish between a \type{conj} and a \type{dnf}.  We
do not consider this implementation to be in the spirit of a primitive
\type{case} expression, because sufficiently complicated
specializations could require non-constant time to execute a
destructor function.  Instead, the first function corresponds to the
least upper bound of \type{conj * dnf} and \type{dnf * conj} in the
upper semi-lattice (i.e., the subtyping hierarchy) induced by the
specializations.  The loss of precision in the resulting type
corresponds to the fact that no specialization exactly corresponds to
the union of the \type{conj} and \type{dnf} specializations.  We could
gain some precision by introducing such a specialization, at the cost
of complicating the interface.

In the examples discussed previously, we have restricted ourselves to
monomorphic, first-order datatypes.  
This restriction can be relaxed to allow polymorphic, first-order
datatypes in the obvious manner: the specialization type makes use of
both non-phantom and phantom type variables, the non-phantom type
variables being applied to the representation type.  For example, the
specializations given in Section~\ref{s:intro} that distinguish
between lists of even and odd length would induce the following
signature:
\begin{codeboxInline}
\begin{verbatim}
  signature EVENODDLIST = sig
    type ('a, 'b) t
    type ('a, 'b) AList = ('a, {list: 'b}) t
    type ('a, 'b) AEven = ('a, {even: 'b}) AList
    type ('a, 'b) AOdd = ('a, {odd: 'b}) AList
    ...
  end
\end{verbatim}
\end{codeboxInline}
and imlementation:
\begin{codeboxInline}
\begin{verbatim}
  structure EvenOddList :> EVENODDLIST = struct
    structure Rep = struct
      datatype 'a t = Nil | Cons of 'a * 'a t
     ...
    end
    type ('a, 'b) t = 'a Rep.t
    type ('a, 'b) AList = ('a, {list: 'b}) t
    type ('a, 'b) AEven = ('a, {even: 'b}) AList
    type ('a, 'b) AOdd = ('a, {odd: 'b}) AList
    ...
  end
\end{verbatim}
\end{codeboxInline}

We may also extend the technique to higher-order
datatypes, although this extension requires the specialization
subtyping hierarchy to induce a full lattice of types (rather than an
upper semi-lattice), due to the contravariance of function arguments.
There are additional restrictions on the subtyping relation at
function types; see our previous work for more
details~\cite{r:fluet02}.

One final limitation of the procedure described here is that it 
applies to a single datatype.  At times, it may be desirable to consider
specializations of one datatype in terms of the specializations of
another datatype.  The technique described in this paper can be
extended to handle this situation by processing all of the specialized
datatypes simultaneously.  Although this decreases the modularity of a
project, this is required in order to define each representation type in terms
of other representation types, which otherwise would be hidden by the
opaque signatures.

It is worth pointing out that some limitations of our implementation
technique are due to the encodings of the subtyping hierarchy induced
by the specializations.  So long as the encodings respect the
hierarchy~\cite{r:fluet02}, the techniques described in this paper are
completely agnostic to the specifics of the encoding.  However, if the
encodings of the subtyping hierarchy possess properties beyond
respecting the hierarchy, these properties may be used to provide a
more flexible implementation of specializations.  The following
example should give a flavor of the kind of flexibility we have in
mind.

\begin{figure}
\hrule
\center{\subfigure[\label{f:bits}The \expr{BITS} signature]{%
\begin{minipage}[t]{0.45\textwidth}
\codeboxInput{code/bits/bits.sig}%
\end{minipage}
}\hspace{\fill}%
\subfigure[\label{f:bits-alt}An alternative \expr{BITS} signature]{%
\begin{minipage}[t]{0.50\textwidth}
\codeboxInput{code/bits/bits-alt.sig}%
\end{minipage}
}}%
\hrule
\caption{}
\end{figure}

Consider a datatype of bit strings, and specializations that capture
the parity of bit strings:
\begin{codeboxInline}
\begin{verbatim}
  datatype bits   = Nil | Zero of bits | One of bits
    withspec even = Nil | Zero of even | One of odd
         and odd  = Zero of odd | One of even
\end{verbatim}
\end{codeboxInline}
Following the approach described in this paper, it is straightforward
to derive an interface $\expr{BITS}$ to the specializations; see
Figure~\ref{f:bits}.  An analysis of the specializations and
implementation reveals that we could safely define a single
\expr{Zero} constructor that applies to all specializations, with type
\type{'a t -> 'a t}. Unfortunately, we cannot similarly define a
single \expr{One} constructor.

However, if we choose the encoding of the subtyping hierarchy
carefully, we can in fact come up with an alternative interface
$\expr{BITS}$ to bit strings and their specializations that allows for
the definition of a single \expr{one} constructor; see
Figure~\ref{f:bits-alt}.  One can verify that the concrete and
abstract encodings respect the induced subtyping hierachy.  However,
they also satisfy a symmetry that makes it possible to write single
instances of the \expr{Zero} and \expr{One} constructors that apply to
all specializations.  In particular, note that the type of the
\expr{One} function makes it explicit that the parity of the bit
string is flipped.

\minisection{Formalizing Results} 
In Section~\ref{s:spec-pt}, we noted that a fully elaborated example
of our specialization technique is more instructive than a formal
translation.  However, it is worth considering how we could formalize 
the technique, and what results we would obtain.
First, we would define a core calculus that captures the essence of
specializations; such a calculus would include a primitive
specialization type with its specializations, constructors into and
destructors from the specializations, and a type system with the
induced subtying relation.  We would prove the type safety of the core
calculus, a consequence of which is that a well-typed program may
never attempt to apply a destructor to an inappropriate
specialization.  We would next define a type-directed translation of
this core calculus into SML.  Finally, we would show that this
translation preserves both the type and the meaning of expressions,
the latter by a simulation argument whereby the runtime behavior of a
well-typed core calculus program is simulated by the translated SML
program.
Since the translation preserves meaning, the translated program will
never raise the \expr{Match} exception---that runtime behavior has no
analogue in the core calculus. 
In fact, the type-preservation result is a special case of
the more general result stating that phantom types may
be used to capture a particular notion of subtyping~\cite{r:fluet02},
where we take the specialization constructors and destructors as
primitive operations.  Under the assumption that the operational
behavior of the constructors and destructors is sound with respect to
their assigned specialization types, we
obtain a result that ensures the absence of runtime exceptions arising
from the use of the primitive operations.  What the current work adds
is the fact that, for datatype specializations, we need not take the
constructors and destructors as primitive operations in the
translation; instead, we may directly encode them in SML.

\section{Conclusion}
\label{s:concl}

We applied two programming techniques, the phantom-types
technique~\cite{r:fluet02} and the recursion-schemes
technique~\cite{r:wang03}, to the problem of capturing structural
invariants in user-defined datatypes. By modeling an abstract
datatype as a collection of constructor, destructor, and coercion
functions, we can define an implementation of datatype specializations
using the techniques developed in this paper.  We further described
methods by which the clumsy destructor functions can be replaced by
familiar pattern matching by injecting into and projecting from
datatypes inspired by recursion schemes.

\begin{sloppypar}
We have collected a set of interesting examples of datatype
specializations at
\url{http://www.cs.cornell.edu/People/fluet/specializations}, all
using the techniques of this paper.  We have also begun work on a tool
to mechanically generate an implementation from the concise
\expr{datatype}/\expr{withspec} declaration.  A proof-of-concept
version of the tool may be obtained at the same website.
\end{sloppypar}

\minisection{Acknowledgments} 
Thanks to Simon Peyton Jones who provided feedback on the application
of phantom types to datatype specializations.
The third anonymous reviewer suggested using substructures as an
organizing principle, rather than ad-hoc name mangling.

\bibliographystyle{entcs}

\end{document}